# The latest results from the Global mm-VLBI Array

**Jeffrey A. Hodgson[1,a], Thomas P. Krichbaum[a], Alan P. Marscher[b], Svetlana G. Jorstad[b], Ivan Marti-Vidal[c], Michael Bremer[d], Michael Lindqvist[c], Pablo de Vicente[e], Anton Zensus[a]**

[a]*Max-Planck-Institut für Radioastronomie, Auf dem Hügel 69, Bonn, Germany*
[b]*Institute for Astrophysical Research, Boston University, 725 Commonwealth Ave, Boston, MA, USA*
[c]*Onsala Space Observatory, Observatorievägen 90, Onsala, Sweden*
[d]*Institut de RadioAstronomie Millimétrique, 300 rue de la Piscine, Grenoble, France*
[e]*Observatorio Astronómico Nacional, Apartado de Correos 112 , Alcalá de Henares, Spain*

E-mail: jhodgson@mpifr-bonn.mpg.de

The Global mm-VLBI Array (GMVA) is the highest angular resolution imaging interferometer currently available as a common user facility. It is capable of angular resolutions on the order of 40 microarcseconds. Currently 14 stations in the United States and Europe participate in global 3 mm VLBI observations. The GMVA is used for continuum and spectroscopic imaging, probing the central regions of active galaxies and the origin of jets as these regions are typically not observable at longer wavelengths due to synchrotron self-absorption. In early 2012, fringes were detected to the three stations of the Korean VLBI Network (KVN), opening the possibility of extending the baseline coverage of the VLBI array to the East. In these proceedings, we will present recent images from a monitoring program of gamma-ray blazars using the GMVA, including the sources 3C454.3 and 0235+164, and an update of its current status and abilities.

---

[1] Speaker

# 1. Introduction

The Global mm-VLBI Array (GMVA)[2] is an array currently consisting of 13 stations spread over the United States and Europe with Yebes also now regularly participating in a best effort capacity. It comprises six European stations (Ef, On, Pv, PdB, Mh and Yb since 2011, see section 2) and the 8 stations of the Very Long Baseline Array (VLBA) equipped with 3 mm receivers. An interferometer has its resolution defined by its operating frequency and the length of its longest baseline, so to increase its resolution, one can either increase the baseline length or the operating frequency. The GMVA achieves its high angular resolution by doing both. The station characteristics are summarised in table 1. An angular resolution of up to ~40 microarcseconds is achieved, though ~50-70 microarcseconds is more typical. 3 mm VLBI will complement the scientific output and interpretation of future space VLBI observations (e.g. RadioAstron at 5 and 22 GHz) and of planned 1 mm VLBI (Event Horizon Telescope).

# 2. Technical description and current status

In October 2011, the 40 m dish in Yebes, Spain for the first time successfully participated in observations, albeit with only one (LCP) polarisation. At 512 Mbps, the 7 σ detection limit on European baselines to Yebes is ~0.1 Jy and ~0.3 Jy to the VLBA. In the May 2012 observations, the Korean VLBI network (KVN) successfully completed a fringe test with fringes detected between the three stations of the KVN and the more sensitive European stations with SNR up to 20. The IRAM interferometer, since being phased, has become a very important part of the GMVA, considerably increasing the sensitivity of the array.

Table 1 – Parameters of current and future GMVA stations

| *Station* | *Location* | *Effective Diameter (m)* | *SEFD (Jy)* |
|---|---|---|---|
| Effelsberg | Germany | 80 | 1000 |
| Plateau de Bure | France | 34 | 500 |
| Pico Veleta | Spain | 30 | 700 |
| Onsala | Sweden | 20 | 5500 |
| Metsähovi | Finland | 14 | 17500 |
| VLBA (x8) | USA | 25 | 2000 |
| Yebes | Spain | 40 | 1700[#] |
| GBT* | USA | 100 | 170 |
| ALMA* | Chile | 85 | 70 |

**NOTE:** *Stations not yet available, but planned. In the future, the Sardinia Radio Telescope (SRT), the Large Millimeter Telescope (LMT) and CARMA may join 3 mm VLBI observations. The Korean VLBI Network (KVN) has successfully participated in test observations with the GMVA. [#]Adjustment of surface-rms in progress (holography).

---

[2] http://www.mpifr-bonn.mpg.de/div/vlbi/globalmm/

The GMVA observes with Mark 5 VLBI recorders at a recording rate of 512 Mbit/s, with a 128 MHz bandwidth. This will shortly be upgraded to 2 Gbit/s recording, effectively doubling sensitivity. Currently, the array is capable of a 7 σ detection limit (in ideal conditions, 10 s integration time) of ~30 mJy. In the future, the Green Bank Telescope (GBT) and the phased ALMA shall be able to join 3 mm VLBI. With the addition of 2 Gbit/s recording and the GBT, 7 σ detection limits lower to ~20 mJy and with phased ALMA to 8 mJy. The GMVA is available for observations in two runs per year in the Spring and Autumn and is open to proposals from the entire astronomical community. The GMVA can also observe in full (LCP and RCP) polarisation, however this still remains challenging to achieve and is described in [6]. Nevertheless, polarisation images have been produced at 3 mm. Data are typically correlated on the DiFX correlator at the MPIfR in Bonn, Germany.

## 3. AGN Science with the GMVA

The study of active galactic nuclei (AGN) is one of the most active areas of research in astronomy today. We are attempting to understand the processes of jet launching, jet kinematics and acceleration, the physical processes of high energy particle production (e.g. gamma rays), magnetic fields and the exact nature of the relationship between the black hole and the base of the jet. High resolution imaging is a key tool in our quest to answer these questions. Contributions from [1], [3] and [8] demonstrate the scientific capabilities of the GMVA on NGC 1052 and 3C 111.

High resolution GMVA observations allow the inner jet and gamma-ray emitting regions to be imaged with unprecedented resolution. When combined with VLBI images at longer wavelengths (e.g. 15 GHz MOJAVE, see [7] or 43 GHz Boston University, see [5]), the position and spectra of VLBI components such as the core and inner jet can be determined. For example, 3C 454.3 [Figures 1,2,3] is one of the brightest and most variable known blazars with broadband emission from radio to gamma-rays. 3C 454.3 is a good example of a source that benefits from being studied at high resolution [2]. At 3 mm, it exhibits complex resolved structure, showing accelerating superluminal components at differing velocities, curved trajectories and large apparent position angle swings in the core region. The data shown here are used to determine the physical properties near the origin of the jet and in the regions where gamma-rays are produced. The images shown here are compared with Boston University 43 GHz images, see [5], where there is a bright downstream jet component. In the GMVA images, superluminal components can be seen being ejected and passing what could be standing shocks.

High resolution spectral decompositions are another method that we can use to investigate the physics of AGN. Producing spectral index maps allows us to probe the spectral properties and magnetic field of the core and inner jet components at sub-parsec scales [4]. Since the synchrotron self absorbed spectrum can be described with the angular size, spectral turnover frequency and flux density, we can determine the location and limit the frequency of the spectral turnover in the jet. This provides us information about the magnetic field strength and geometry of the jet. We can determine jet speed from VLBI and combine it with the Doppler factor from radio and gamma-ray variability to derive the viewing angle. We then are able to measure the jet geometry and constrain the areas in which gamma-rays are produced.

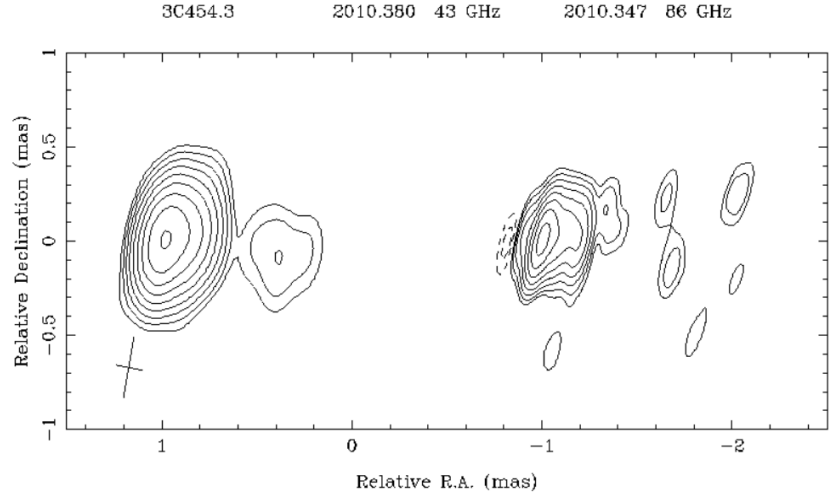

**Figure 1** – 43 GHz Boston University image of 3C 454.3 (left, peak: ~21.1 Jy/beam) aligned at 1 mas with a beam of 0.28 x 0.12 mas, and GMVA image (preliminary, right) aligned at -1 mas with a beam of 0.25 x 0.07 mas. The elongation of the core in the easterly direction is clearly resolved in the GMVA image. Contours start at 0.08 mJy in steps of two. Both images are from May 2010.

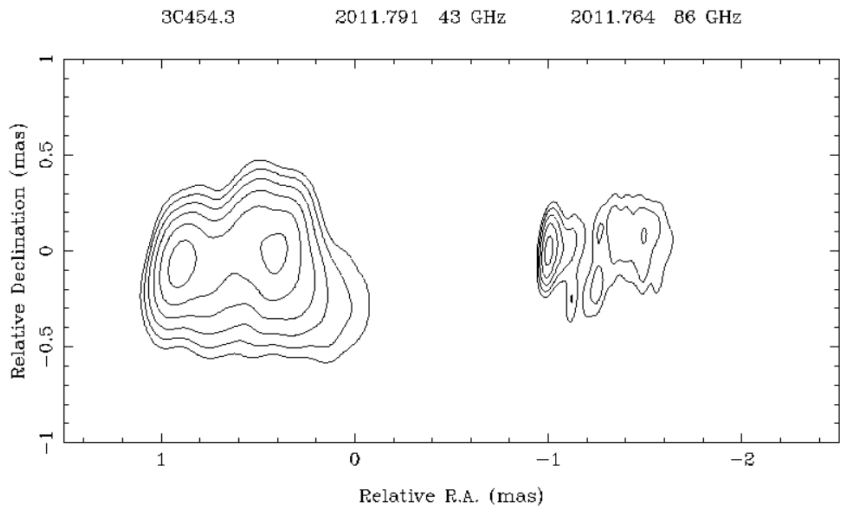

**Figure 2** – 43 GHz Boston University image of 3C 454.3 (left, peak ~1.8 Jy/beam) aligned at 1 mas with a beam of 0.28 x 0.12 mas, and GMVA image (preliminary, right, peak ~0.87 Jy/beam) aligned at -1 mas with a beam of 0.26 x 0.06 mas. However, the downstream component is very bright in the 43 GHz image, but in the GMVA images, there is complicated structure and what appears to be superluminal components approaching what could be a standing shock. Contours start at 0.02 mJy in steps of two. The component south-east of the core is suspicious and must be model fitted to be confirmed. Both images are from Oct 2011.

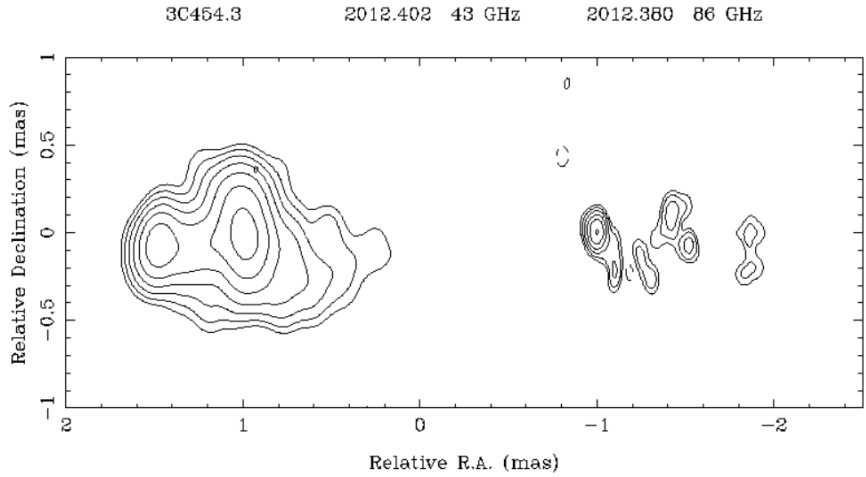

**Figure 3** – 43 GHz Boston University image of 3C 454.3 (left, peak ~0.9 Jy/beam) aligned at 1 mas with a beam of 0.28 x 0.12 mas, and GMVA image (preliminary, right, peak ~1.1 Jy/beam) aligned at -1 mas with a beam of 0.20 x 0.09 mas. Unlike the observations of 2010 and 2011, the elongation of the core is now in a south-easterly direction and unresolved at 43 GHz. As in the 2011 images, the downstream component is bright in the 43 GHz image. However, it appears that the superluminal component has passed what could be a standing shock. Contours start at 0.01 mJy in steps of two. Both images are from May 2012.

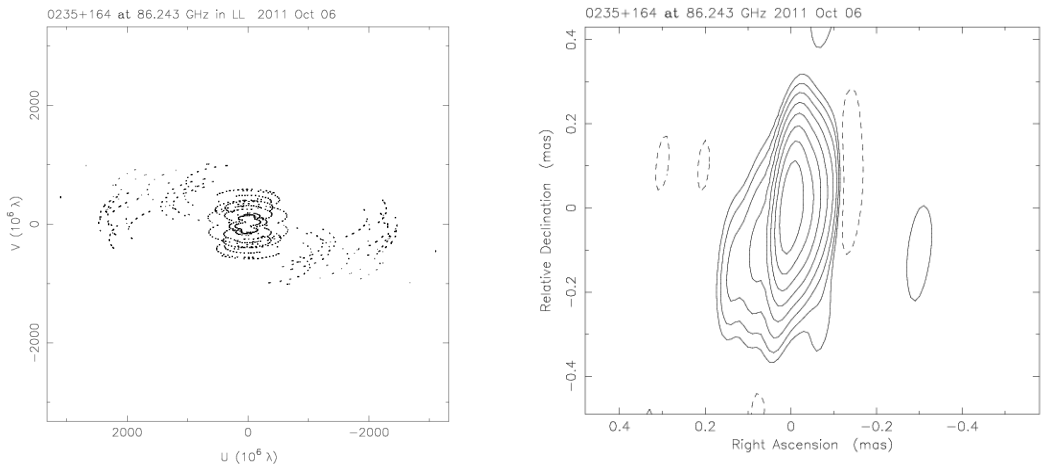

**Figure 4** – LEFT: Typical UV coverage of a snapshot observation using the GMVA. RIGHT: October 2011 3 mm VLBI image of 0235+164 (Peak: 1.07 Jy/beam; Beam: 0.22 x 0.052 mas; Contours: 0.04 mJy in steps of two). Structure resolved at 3 mm, which is unresolved at lower frequencies.

## 4. Future Outlook

Given that at higher resolutions, the position of resolved components may change considerably between 6 month observations, especially those with curved trajectories, it would be highly desirable to have more frequent observations with more flexible scheduling. The source 0235+164 [Fig 4, right], which is one of the most variable gamma-ray blazars, is unresolved at 7 mm and lower frequencies. At 3 mm, possible signs of structure can be seen in what is likely to be one of the most highly superluminal sources. More frequent and flexible observations would allow these motions to be reliably detected. The addition of the KVN offers two fold improvements. The first is the possibility of high resolution baselines when there is no mutual visibility between Europe and the USA. The second is that it provides short baseline spacings that are important in aiding calibration and the detection of extended structure. With the addition of a phased ALMA and the GBT, the array will be considerably more sensitive, hence vastly expanding the number of sources that are observable. In the future, the LMT in Mexico, the SRT in Italy and CARMA could also join the GMVA. We also note that Mopra and the ATCA in Australia are 3 mm capable. With the combination of these antennas, mm VLBI would be comparable in sensitivity with cm wavelength VLBI – at far higher angular resolution. The GMVA will also complement planned 1 mm Event Horizon Telescope observations. The EHT has only very limited UV coverage and sensitivity and hence its imaging capabilities are severely limited. High quality 3 mm observations will be necessary to aid the interpretation of EHT experiments and model fitting. The matched resolution of the GMVA at 86 GHz and RadioAstron at 22 GHz allows for the spectra to be determined of the innermost 20-40 microarcseconds of AGN, allowing for the determination of magnetic field strengths.